\documentclass[preprint]{vgtc}               

\ifpdf
  \pdfoutput=1\relax                   
  \usepackage{graphicx}                
  \DeclareGraphicsExtensions{.pdf,.png,.jpg,.jpeg} 
\else
  \usepackage{graphicx}                
  \DeclareGraphicsExtensions{.eps}     
\fi%

\graphicspath{{figures/}{pictures/}{images/}{./}} 

\usepackage{microtype}                 
\PassOptionsToPackage{warn}{textcomp}  
\usepackage{textcomp}                  
\usepackage{mathptmx}                  
\usepackage{times}                     
\usepackage{cite}                      
\usepackage{tabu}                      
\usepackage{booktabs}                  

\usepackage{listings}
\usepackage[dvipsnames]{xcolor}
\usepackage[frozencache=true,cachedir=minted-cache]{minted}
\usepackage{enumitem}
\usepackage{url}
\usepackage[title]{appendix}
\usepackage{subfig}
\usepackage{rotating}
\usepackage{tikz}
\usepackage{newfloat}
\usepackage{caption}
\usepackage{balance}
\usepackage{amssymb} 
\usepackage[breaklinks=true]{hyperref}
\usepackage[capitalize]{cleveref}

\definecolor{codegreen}{rgb}{0,0.6,0}
\definecolor{codeblue}{rgb}{0,0,0.6}
\definecolor{codegray}{rgb}{0.5,0.5,0.5}
\definecolor{codepurple}{rgb}{0.58,0,0.82}

\usepackage{ascii}

\lstdefinestyle{mystyle}{
    language=Python,
    commentstyle=\color{codegray},
    keywordstyle=\color{codegreen},
    numberstyle=\tiny\color{codeblue},
    stringstyle=\color{codepurple}\scriptsize\asciifamily,
    basicstyle=\scriptsize\asciifamily,
    breaklines=true,                 
    captionpos=b,                    
    keepspaces=true,                 
    numbers=left,                    
    showstringspaces=false,
    tabsize=2,
    morekeywords={True, False},
    basewidth={.55em}
}
\onlineid{1050}

\vgtcpapertype{Empirical studies for evaluation}

\vgtcinsertpkg

\preprinttext{Submitted to IEEE VIS.}

\title{Data Point Selection for Line Chart Visualization: \\ Methodological Assessment and Evidence-Based Guidelines}

\newcommand*\samethanks[1][\value{footnote}]{\footnotemark[#1]}

\author{
    Jonas Van Der Donckt 
    \thanks{contributed equally} 
    \thanks{e-mail: jonvdrdo(dot)(lastname)(at)ugent(dot)be} 
    \and Jeroen Van Der Donckt  
    \samethanks[1]
    \and Michael Rademaker  
    \and Sofie Van Hoecke  
}
\affiliation{\scriptsize IDLab, Ghent University - imec, Belgium}

\teaser{
  \centering
  \includegraphics[width=\linewidth]{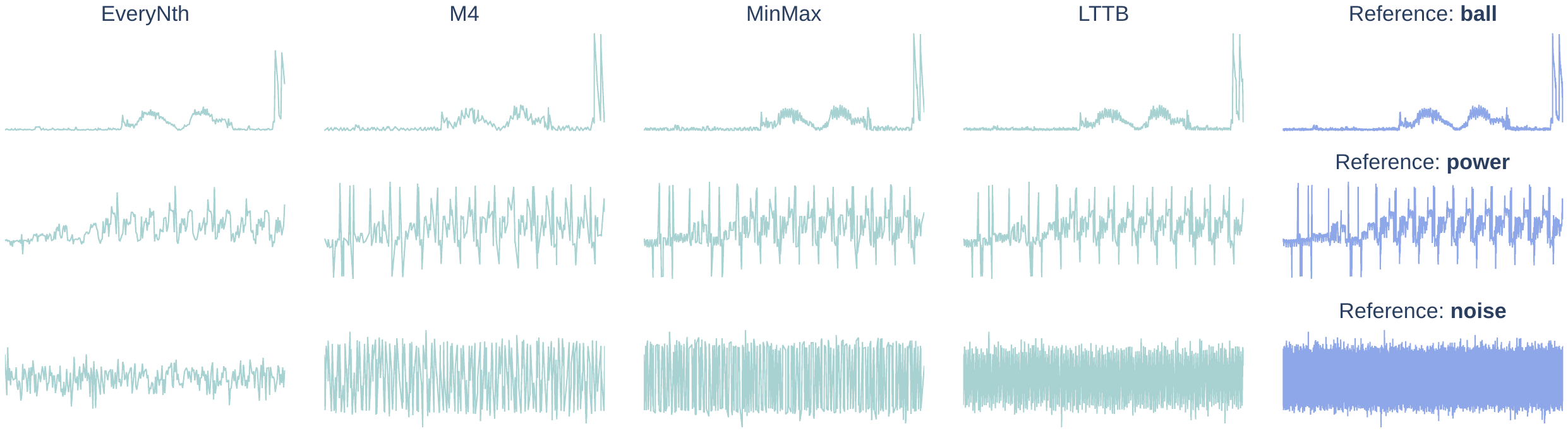}

  \caption{
    Visual analysis of four time series data point selection algorithms. 
    Each row corresponds to a unique template, with the rightmost column presenting a reference image, constructed from 200k data points. The four adjacent columns demonstrate the varying visual representations, each generated by a data point selection algorithm, all employing an equal number of output data points ($n_{out} = 300$). Visualization properties used include: Plotly as toolkit, line-width set to 2, and anti-aliasing enabled.
    A \href{https://github.com/predict-idlab/ts-datapoint-selection-vis/blob/main/gifs/teaser_n_out.gif}{GIF} further demonstrates the above visualization with varying values for $n_{out}$.
  }
  \label{fig:teaser}
}

\abstract{%
  Time series visualization plays a crucial role in identifying patterns and extracting insights across various domains. However, as datasets continue to grow in size, visualizing them effectively becomes challenging. Downsampling, which involves data aggregation or selection, is a well-established approach to overcome this challenge.
  This work focuses on data selection algorithms, which accomplish downsampling by selecting values from the original time series. Despite their widespread adoption in visualization platforms and time series databases, there is limited literature on the evaluation of these techniques.
  To address this, we propose an extensive metrics-based evaluation methodology. Our methodology analyzes visual representativeness by assessing how well a downsampled time series line chart visually approximates the original data. 
  Moreover, our methodology includes a novel concept called "visual stability", which captures visual changes when updating (streaming) or interacting with the visualization (panning and zooming).
  We evaluated four data point selection algorithms across three open-source visualization toolkits using our proposed methodology, considering various figure-drawing properties.
  Following the analysis of our findings, we formulated a set of evidence-based guidelines for line chart visualization at scale with downsampling. 
  To promote reproducibility and enable the qualitative evaluation of new advancements in time series data point selection, we have made our methodology and results openly accessible. The proposed evaluation methodology, along with the obtained insights from this study, establishes a foundation for future research in this domain.
}

\keywords{Time series, Line charts, Downsampling, Data aggregation, Perception, Evaluation, LTTB, M4, MinMax.}

\nocopyrightspace

\begin{document}



\maketitle

\section{Introduction}
Time series data are commonly found in various fields, such as healthcare, finance, and engineering. This type of data often comprises complex information that cannot be fully understood through summary statistics alone, making visualization an essential tool for identifying patterns and extracting insights. Among the various visualization methods, line charts have proven to be particularly effective in analyzing time series in diverse applications~\cite{aigner_visualizing_2007}.

The size of time series data has been increasing continuously, often comprising millions to billions of data points. However, handling and visualizing such large datasets poses a significant challenge. As a result, over the last decade, there has been a surge in the development of time series databases, such as InfluxDB, TimescaleDB, OpenTSDB, and Prometheus, designed to handle and persist large-scale time series data. 
To efficiently visualize the data, downsampling has emerged as an established technique that involves aggregating or selecting a representative subset of the time series and interpolating between the downsampled points~\cite{kwon_sampling_2017, agrawal_challenges_2015, aigner_visualizing_2007}. Downsampling reduces the number of data points while aiming to preserve the overall shape of the time series, this way minimizing network latency and improving rendering time. It has therefore become an essential component in many widely adopted time series database solutions.

This work specifically focuses on evaluating data point selection algorithms, also referred to as value preserving aggregation~\cite{jugel_vdda_2016, jugel_m4_2014}, which select data points from the original time series to downsample the data. 
Bearing computational efficiency in mind, we pinpointed the most well-established data point selection algorithms in practice, comprising EveryNth (selecting every $n^{th}$ data point), MinMax, Largest-Triangle-Three-Buckets (LTTB)~\cite{steinarsson_downsampling_nodate}, and M4~\cite{jugel_m4_2014}.
These algorithms are broadly adopted in industry, with examples such as Uber incorporating LTTB as a downsampling function in their M3 metrics platform~\cite{uber_raskin_aggarwal_2018} and TimeScaleDB offering LTTB as a server-side hyperfunction~\cite{timescaleblog_paganini_2023}. Despite their widespread use, a profound evaluation of these algorithms remains under-studied due to the absence of a one-size-fits-all metric, the complex interplay of visualization components (e.g., line width, line shape, toolkit), and the lack of attention to visual stability during limited interactions such as small zooms or panning.

To address these challenges, this work proposes an extensive metrics-based methodology for evaluating the visual quality of data point selection algorithms, focusing on two key aspects: visual representativeness and visual stability. Visual representativeness assesses the extent to which a downsampled time series line chart visually approximates the original data by employing image-based metrics, as such capturing toolkit and line drawing characteristics. Visual stability, a novel concept that we introduce, quantifies the perceptual coherence between two slightly different aggregations, which is particularly relevant when updating visualizations gradually, such as during data streaming, panning, or small zooms. These interactions have been demonstrated to be the most effective means for visually exploring time series data~\cite{shneiderman_eyes_nodate, van2022plotly, walker_timenotes_2016}, making visual stability an essential aspect of time series downsampling for visualization.
Additionally, our methodology incorporates various time series templates and advocates for using the number of selected data points as a dependent variable, enabling insightful comparisons with regard to data efficiency.


Applying the proposed methodology, we evaluate the four aforementioned data point selection algorithms across three visualization toolkits, considering various figure drawing properties, such as line width and anti-aliasing. Based on the results, we derive a set of guidelines for line chart visualization at scale with downsampling. 

To summarize, the contributions of this paper are threefold.
First, we present a robust and extensive metrics-based evaluation methodology for data point selection algorithms, quantifying visual representativeness and a novel concept called "visual stability." 
Second, employing the proposed methodology, we assess four prominent downsampling algorithms, taking into account multiple visualization toolkits and a variety of figure-drawing properties. The insights gained from this analysis are summarized into a set of guidelines for large-scale time series visualization with downsampling.
Lastly, we provide a high-quality open-source implementation of the proposed framework along with the obtained results, ensuring reproducibility and paving the way for the analysis and improvement of novel downsampling algorithms, available at \href{https://github.com/predict-idlab/ts-datapoint-selection-vis}{github.com/predict-idlab/ts-datapoint-selection-vis}.
\section{Related Work}
This section provides a comprehensive overview of related work in the field. We first examine time series visualization techniques, and then discuss high-level methods for data aggregation. Specifically, we focus on value preserving data aggregation algorithms, considering both their practical applicability and existing research in this area. By doing so, we aim to offer a clear understanding of the current state-of-the-art and identify the gaps that our work seeks to address.

\subsection{Time Series Visualization}\label{sec:related-work-time-series-vis}
Visualization plays a crucial role in understanding time series data, especially given the complexity of this data modality. After all, the human eye has frequently been advocated as the ultimate data mining tool~\cite{lin_visualizing_2005}. Simple visualizations, such as line-based charts, have been found to be sufficient for most time series analysis tasks, as opposed to more complex or specific approaches~\cite{aigner_visualizing_2007}. Hence, this work will focus on the representativeness of line chart visualizations.

Interactive visualization techniques are essential for exploring large amounts of time series data. Shneiderman et al. proposed the visual information seeking mantra, which emphasizes the importance of exploring an overview of the data, followed by zooming and filtering before accessing details-on-demand~\cite{shneiderman_eyes_nodate}. Moreover, Walker et al. stress that interactive line chart visualizations improve the efficiency of performing visualization-oriented tasks~\cite{walker_timenotes_2016}.

In previous work, we contributed \texttt{plotly-resampler}, which is a wrapper for the Plotly visualization library's Python bindings that adds zoom-plot functionality through a back-end that operates on graph interaction callbacks~\cite{van2022plotly}. Two major advantages of \texttt{plotly-resampler} over other existing tools are that it does not require a DBMS, and it enables trace-level specification of the data aggregation configuration. 
The obtained results from this study can be directly transferred to \texttt{plotly-resampler}.

\subsection{Data Aggregation for Line Chart Visualization}
Data aggregation techniques have become crucial for addressing big data and visual analytics challenges~\cite{kwon_sampling_2017, gorodov_analytical_2013, bikakis_big_2018}. These methods tackle several challenges that arise when visualizing large datasets. In particular, front-end rendering and responsiveness slow down significantly when more data points are transferred and portrayed, making data aggregation essential. 
Furthermore, a common side effect of visualizing large amounts of data, is over-plotting, where multiple data points overlap due to limited visualization canvas size~\cite{datashaderpitfalls, walker_timenotes_2016}. Data aggregation leverages this effect by potentially omitting some of the overplotted data points. 

There are two different approaches to performing data aggregation for visualization: downsampling and density-wise aggregation. 
Downsampling outputs a lower-dimensional representative dataset, while density-wise aggregation employs shared density color coding to render a data image. Downsampling can be further divided into value preserving and characteristic data aggregation. Below, we discuss density-wise and characteristic data aggregation, while \cref{sec:value_preserving_agg} delves into value preserving aggregation, which is the focus of this work.


\subsubsection{Density-wise Aggregation}
Density-wise data aggregation involves using a shared density color coding to render an image of the data~\cite{breddels_interactive_2016}. While this technique is effective for visualizing single, large data modalities such as maps and point clouds, it poses challenges when there are multiple distinct modalities in the data. Each modality would then require a distinguishable color density coding, often leading to over-plotting. However, time series data typically consists of distinct modalities, making this technique less preferable. Moreover, converting data into images restricts differentiation within subplots and interactivity, such as toggling time series traces, further reducing its suitability for time series visualization.

Datashader is the only known implementation of density-wise data aggregation that works for time series data (i.e., line charts)~\cite{datashader}. 



\subsubsection{Characteristic Aggregation}
Characteristic data aggregation reduces data by highlighting properties or trends, using methods like mean, median, and smoothing. As these techniques' objectives cannot be objectively quantified, representativeness evaluation requires user studies.

The Control project, introduced in the 1990s, was among the first to utilize sampling for scalable aggregation queries, such as SUM, COUNT, and AVG~\cite{hellerstein1999control}. This approach enabled prompt visualization of approximate query results while allowing users to specify confidence intervals. By communicating result uncertainty, users could trust the visualization~\cite{kwon_sampling_2017}.

Alternatively, Rong et al. presented ASAP (Automatic Smoothing for Attention Prioritization)~\cite{rong_asap_2017}, which smooths time series by adaptively optimizing noise reduction and trend retention, directing users' attention towards significant deviations.

\subsection{Value Preserving Algorithms}\label{sec:value_preserving_agg}
Value preserving data aggregation selects data points from the original time series while aiming to preserve its overall shape. To do so, such techniques leverage interpolation between the selected data points. 

Unlike characteristic and density-wise aggregation techniques, the objective of value preserving aggregation is to approximate the ground truth image. The ground truth image, also referred to as the template or reference image, is the visualization that includes all the data points~\cite{jugel_m4_2014, steinarsson_downsampling_nodate}. Thus, the visual representativity of value preserving methods can be evaluated using image quality assessment metrics~\cite{zhai2020perceptualIQA}. Note that the ground truth image may suffer from overplotting.

\subsubsection{Practical Applicability}
For real-world, large-scale time series data, downsampling algorithms must meet specific computational requirements. These include: (i) at most $O(N)$ runtime complexity, i.e., a single iteration over the data; (ii) avoidance of additional data structures scaling with data size, as maintaining and updating such structures is computationally infeasible; and (iii) parallelizable to alleviate linear scaling by distributing data partitions across multiple threads or cores.

\begin{table}[]
\caption{Overview of data point selection algorithms, where $N$ denotes the time series length, and $n_{out}$ represents the aggregation output size.}
\label{tab:aggregators_overview_}
\resizebox{\columnwidth}{!}{%
    \begin{tabular}{@{}rccccc@{}}
    \toprule
    \textbf{} & \textbf{\begin{tabular}[c]{@{}c@{}}(i) \\ Time\end{tabular}} & \textbf{\begin{tabular}[c]{@{}c@{}}(ii) \\ Memory\end{tabular}} & \textbf{\begin{tabular}[c]{@{}c@{}}(iii) \\ Parallelizable\end{tabular}} & \textbf{\begin{tabular}[c]{@{}c@{}}Data points \\ per bucket\end{tabular}} \\ \midrule
    EveryNth & $O(n_{out})$ & $O(1)$ & \checkmark & 1 \\
    MinMax & $O(N)$ & $O(1)$ & \checkmark & 2 \\
    M4 \cite{jugel_m4_2014} & $O(N)$ & $O(1)$ & \checkmark & $\pm$4 \\
    LTOB \cite{steinarsson_downsampling_nodate} & $O(N)$ & $O(1)$ & \checkmark & 1 \\
    LTTB \cite{steinarsson_downsampling_nodate} & $O(N)$ & $O(1)$ & $\times$ & 1 \\
    \midrule
    ModeMedianB \cite{steinarsson_downsampling_nodate} & $O(N)$ & $O(N/n_{out})$ & $\pm$ & 1 \\
    Visval \cite{visvalingam1993line} & $O(N\log{(N)})$ & $\Omega(N)$ & $\pm$ & 1 \\
    DP \cite{douglas1973algorithms} & $O(N \log{(N)})$ & $\Omega(N)$ & $\pm$ & 1 \\
    MinStdErrB \cite{steinarsson_downsampling_nodate} & $O(N^2)$ & $O(N^2)$ & $\times$ & 1 \\
    LLB \cite{steinarsson_downsampling_nodate} & $O(N^2)$ & $O(N^2)$ & $\times$ & 1\\
    \bottomrule
    \vspace{-3mm}
    \end{tabular}
}
\end{table}

\cref{tab:aggregators_overview_} summarizes value preserving algorithms for time series visualization concerning the three outlined requirements. Note that the memory complexity captures additional memory beyond the input and output arrays.
Only the five algorithms above the horizontal line satisfy the proposed requirements. Among these, Largest-Triangle-One-Bucket (LTOB) and Largest-Triangle-Three-Buckets are based on triangular surface maximization. However, since LTOB is a shortsighted variant of LTTB (considering only the two neighboring data points)~\cite{steinarsson_downsampling_nodate} and Gil et al.~\cite{gil2021towards} demonstrated LTTB's superiority over LTOB, we excluded LTOB from this study.

The algorithms below the horizontal line fail to meet the proposed requirements. The table includes two line simplification algorithms: Douglas-Peucker (DP)\cite{douglas1973algorithms} and Visvalingam-Whyatt (Visval)\cite{visvalingam1993line}. Both are unsuitable for large-scale time series visualization due to their $O(N \log{(N)})$ time complexity bounds and their use of $\Omega(N)$ data structures for limiting re-computation. Additionally, the table lists three other algorithms introduced by Steinarsson~\cite{steinarsson_downsampling_nodate} that are deemed unsuitable, with Mode-Median-Bucket exhibiting an undesired $O(N/n_{out})$ memory complexity, and Min-Std-Error-Bucket having $O(N^2)$ runtime and memory complexity.

In the following subsections, we will discuss four value preserving algorithms that are computationally suitable (excluding LTOB) for time series visualization at scale, forming the focus of this study.

\subsubsection{EveryNth}
The EveryNth algorithm, also known as sampling or decimation, is a highly efficient method for value preserving downsampling. This approach selects every $n^{th}$ data point from the input data array, where $n$ is the ratio of the input array length ($N$) to the desired output array length ($n_{out}$).
The algorithm has a time complexity of $O(n_{out})$ and a memory complexity of $O(1)$. Notably, it is the only algorithm among those listed in \cref{tab:aggregators_overview_} that scales linearly with $n_{out}$ instead of $N$. Although easily parallelizable, multi-threading may not offer significant benefits due to the simplicity of the sampling task.



\subsubsection{MinMax}
MinMax aims to preserve the shape of time series data by selecting vertical extrema for each bucket. The algorithm selects the minimum and maximum values within each bin, but does not enforce alternating patterns. This can result in repeated neighboring minima (or maxima), potentially failing to correctly capture underlying alternations. 
MinMax has a time complexity of $O(N)$ and a memory complexity of $O(1)$. It is easily parallelizable, enabling efficient execution on multicore processors.
%


%



\subsubsection{M4}
M4, proposed by Jugel et al.~\cite{jugel_m4_2014}, is a value preserving aggregation method that can achieve pixel-perfect data aggregation. For each bin, it selects two horizontal (first, last) and two vertical (min, max) extrema. Selecting these data points can be done in  linear time complexity ($O(N)$) with $O(1)$ memory complexity. The algorithm is also easily parallelizable. 
Remark that M4 selects more data points per bin than the other algorithms. As such, Bae et al. introduced inter-pixel gradient-based M4 (IGM4)~\cite{bae2017practical}, which removes redundant data points as a post-processing step. This reduces the selected data points by an average of 20\%, which is still considerably more data points per bin than the other downsampling algorithms.

\begin{figure*}[!tb]
 \centering
 \includegraphics[width=\linewidth]{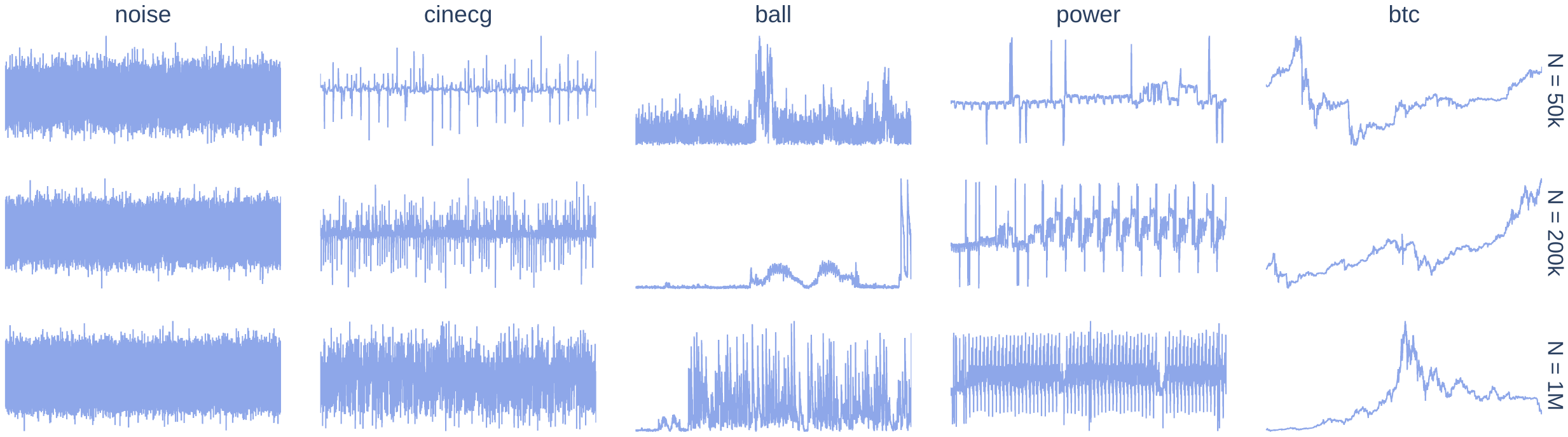}
 \caption{Grid of utilized time series templates. Each column represents a distinct time series, and the rows indicated the template sizes.}
 \vspace*{-3mm}
 \label{fig:template_grid}
\end{figure*}

\subsubsection{Largest-Triangle-Three-Buckets (LTTB)}
Steinarsson proposed several value preserving downsampling algorithms~\cite{steinarsson_downsampling_nodate}. This work focuses on the LTTB algorithm, as the other proposed algorithms either have unfavorable time complexity or have poorer visual representativeness compared to LTTB, see \cref{tab:aggregators_overview_}.
LTTB is based on the concept of effective triangular area, which is often employed in line simplification algorithms. It selects data points in each bucket (i.e., bin) that form the largest triangular surface with the previously selected data point and the next bucket's average value. LTTB has a time complexity of $O(N)$ and a memory complexity of $O(1)$. Although its runtime scales linearly with data size, it has a steeper slope compared to MinMax and M4 due to the increased cost of area calculations. LTTB cannot be parallelized, as it requires sequential pass over the data since the previously selected data point is part of the local surface calculation

\subsection{Evaluating Time Series Data Aggregation}
Several works have made steps toward assessing the visual representativeness of data aggregation for time series visualization. 

Shi et al.~\cite{shi2006performance} proposed two data-based metrics for evaluating cartographic line simplification: displacement measure (the maximal perpendicular distance between the simplified and original line) and shape distortion measure (the average difference in inclination angles between the simplified and original line).

Jugel et al.\cite{jugel_m4_2014} used the Structural Similarity Index Measure (SSIM) as an image-based metric to evaluate and compare their M4 data aggregation algorithm. However, the reliability of SSIM has been questioned by Nilsson et al.\cite{nilsson2020understanding} and Sara et al.~\cite{sara2019image}, who both demonstrated that assessing image quality based on SSIM can lead to incorrect conclusions and does not consistently capture reconstruction errors.

Gil et al.~\cite{gil2021towards} contributed a method for adaptively identifying optimal data point selection algorithms for time series compression, using multiple value preserving algorithms and a data-based error score. However, the approach has limitations: MinMax was not included, only mean absolute error (MAE) was evaluated, and interpolating the downsampled signal to the original signal results in $O(N)$ memory complexity. Furthermore, their results suggest that using LTTB or M4 is only marginally worse than the optimal strategy, questioning the practical applicability of their approach.
To the best of our knowledge, no prior research has assessed the stability of downsampling methods when updating (streaming) or interacting with the visualization (panning and zooming). However, as outlined in \cref{sec:related-work-time-series-vis}, interactivity serves as the most efficient approach for exploring time series visualizations. Consequently, visual stability will be one of the two core aspects of our proposed methodology.

\section{Methodology}
The purpose of this study is to evaluate and compare time series data point selection algorithms for visualization. Specifically, we focus on two aspects: visual representativeness and visual stability.

\textbf{Visual representativeness} refers to the degree to which a downsampled time series graph visually reflects the original rendered data. To assess this, we use three image-based metrics that are each aggregated, i.e., scaled, using an OR-convolution mask. 

\textbf{Visual stability} aims to capture the visual changes that occur when zooming, panning, or streaming, i.e., when re-rendering with a largely overlapping view. To assess visual stability, we use two data-space metrics that focus on residual errors.

To ensure robust and fair comparisons of the downsampling algorithms, our methodology employs multiple time series datasets that are representative of a wide range of time series datasets. Additionally, we advocate using the number of downsampled points as a measure of data efficiency, facilitating cross-algorithm comparisons.

In this section, we first describe the study design, including the time series templates and how data efficiency should be evaluated. Then, we explain the image-space metrics for visual representativeness assessment, followed by an introduction to the data-space metrics for assessing visual stability.


\subsection{Study Design}
Our methodology employs multiple time series templates that encompass a range of time series properties, including periodicity and noisiness. The length of these time series templates is varied, providing a comprehensive representation of different zoom levels of the same data source. Additionally, we recommend using $n_{out}$ as a fair measure for comparing the data efficiency of downsampling algorithms.

One crucial aspect of our study design is the use of a fixed canvas size for all visualizations. In this study, we have chosen for a (800, 250) canvas, in accordance with Jugel et al.~\cite{jugel_m4_2014}. By maintaining a constant canvas size while varying $n_{out}$, we ensure that our comparisons are consistent, enabling us to draw conclusions about the relationship between $n_{out}$ and canvas width. 
Furthermore, since our study focuses solely on univariate time series visualization, which involves a single line (per) chart, we opted to render the line-chart visualizations in grayscale. This decision was based on the fact that there is no need to differentiate between multiple modalities in the same plot. As a result, the resulting visualization images are 2D matrices, with each pixel containing only a single illumination value ($[0, 255]$). 

\subsubsection{Time Series Templates}
To ensure a comprehensive evaluation, our methodology uses multiple time series templates that cover a wide range of time series properties, such as periodicity, stationarity, noisiness, spanning surface of time series on visual canvas, and the presence of high-frequency components. These templates, also referred to as image templates, represent signals of a specific size $N$. The extent of $N$-variation in these templates provides a comprehensive representation of various zoom levels, which is a prominent way of interacting with time series data~\cite{aigner_visualizing_2007}. We selected the sizes $N \in \{ 50K, 200K, 1M \}$ through empirical analysis. These sizes were deemed practical for visualization purposes (i.e., capable of rendering the reference image without crashing), while also providing informative views of the data. 

Our study utilized five distinct time series signals that can be observed in \cref{fig:template_grid}. These include: 
(1) generated Gaussian noise, representing a high-frequency stationary noise; (2) an ECG signal with prominent peaks and periodicity from the UCR time series archive~\cite{dau2019ucr}, used by Franco et al.~\cite{franco2022multivariate}; (3) a positively peaked ball speed signal from the DEBS 2013 dataset~\cite{mutschler2013debs}, used by Jugel et al.~\cite{jugel_m4_2014, jugel_vdda_2016}; (4) a periodic power consumption signal from the DEBS 2012 dataset~\cite{jerzak2012debs} exhibiting prominent peaks, used by Jugel et al.~\cite{jugel_m4_2014, jugel_vdda_2016}; and (5) a non-stationary signal constructed of 1-minute intervals of historical bitcoin data~\cite{kottarathil_btc_data_2022}, in accordance with the stock data used by Steinarsson~\cite{steinarsson_downsampling_nodate}. \cref{fig:template_grid} shows that as the template size $N$ increases for both the ECG and power signals, more peaks and periods become visible. Remark that the ball time series signal displays completely different patterns as $N$ changes.




\subsubsection{Data Efficiency}
When assessing the effectiveness of downsampling algorithms, it is crucial to compare them based on the number of output data points 
$n_{out}$. This number is directly related to the amount of data that is transmitted to and displayed in the front-end, which affects visualization performance. While previous studies have focused on the number of bins used in downsampling~\cite{jugel_m4_2014, jugel_vdda_2016}, these overlook the fact that different algorithms can have varying numbers of data points per bin (as shown in \cref{tab:aggregators_overview_}). Therefore, we suggest using $n_{out}$ as a fair measure for comparing downsampling algorithms.



\subsection{Visual Representativeness}
Visual representativeness evaluates how closely a downsampled time series graph visually approximates the original rendered data, also referred to as the reference visualization. To assess this, the visualization of the downsampled time series is compared with the reference visualization, which corresponds to comparing images.
As such, we use three image-space metrics that capture various aspects of the comparison: MSE, pixel errors (with a margin), and, DSSIM. All three metrics are considered full-reference image quality assessment measures and are classified as spatial domain methods~\cite{zhai2020perceptualIQA}. Our goal is to capture both absolute errors with MSE and pixel errors, as well as perceived quality with (D)SSIM. Although MSE is commonly used as a measure of signal fidelity, it has been criticized for not taking into account certain image signal characteristics and human perception~\cite{wang2009mean}\footnote{Note that PSNR is not relevant for this specific task since the maximum pixel value of the image will always be 255, which results in displaying identical trends as to MSE.}. Therefore, since human perception is highly sensitive to structural information in images, we use a dissimilarity variant of the structural similarity index measure.
By using image-space metrics, we can investigate the impact of the various parameters that affect the visualization. These include, but are not limited to, the rasterization back-end, plotting toolkit, line-drawing style, line width, and anti-aliasing. Of particular interest is that this allows evaluating the default configurations used in visualization toolkits. 

Below, we outline the three image-space metrics for evaluating visual representativeness, followed by a description of the OR-convolution mask which we use to aggregate each of the three per-pixel metric outputs into a single score. 




\subsubsection{Mean Squared Error}
The first metric utilized to assess visual representativeness is the Mean Squared Error (MSE). For each pixel, the squared error is calculated, which is then aggregated into a single value for all pixels that are within the OR-conv mask.
This measure captures absolute pixel errors, but unlike mean absolute error, MSE penalizes less for subtle shading errors resulting from anti-aliasing.

\subsubsection{Pixel Errors}
The second metric for assessing visual representativeness quantifies pixel errors by binarizing them. 
A pixel is classified as an error (assigned a value of 1) when its value in the downsampled visualization differs from the reference visualization.
To aggregate the pixel errors into a single measure, we first calculate the total number of pixel errors and then normalize it by dividing with the OR-conv mask's area.
Unlike MSE, this metric equally penalizes even the slightest differences in pixel values. However, this equal penalization can make the metric less interpretable for anti-aliased figures, as it penalizes subtle shading differences that are not noticeable to human perception. To address this, we introduce a margin of 20 on the pixel differences to include only visible errors. This metric is called Pixel Error Margin 20 (PEM\_20).


\subsubsection{(D)SSIM}\label{sec:dssim}

As a third metric, we use the Structural Dissimilarity ($DSSIM$), a scaled version of the Structure Similarity Index Measure ($SSIM$), which was also employed by Jugel et al.~\cite{jugel_m4_2014, jugel_vdda_2016}. $SSIM$ is a commonly used metric for assessing perceived quality~\cite{nilsson2020understanding}, producing a similarity score ranging from -1 to 1. A score of 1 indicates perfect similarity, 0 indicates no similarity, and -1 indicates perfect anti-correlation. As such, the dissimilarity between images $V_1$ and $V_2$ is defined as:

$$ DSSIM(V_1, V_2) = \frac{1 - SSIM(V_1, V_2)}{2} $$

Just as for the previous metrics, the output of $(D)SSIM$ is a matrix that contains a value for each pixel. Jugel et al. aggregated the per-pixel values into a single measure by applying a global mean\footnote{Jugel et al. confirmed to the authors that they performed a data inversion on the $DSSIM$ in~\cite{jugel_vdda_2016, jugel_m4_2014}, essentially casting the $DSSIM$ from $[0: 1]$ to $[1: 0]$.}. In this work, however, an OR-conv mask is applied to mean-aggregate only $DSSIM$ values where line charts are shown, see \cref{sec:conv_mask}.



\subsubsection{OR-conv Masking}\label{sec:conv_mask}
We introduce a novel technique called "OR-conv masking" to aggregate per-pixel metrics in a more targeted manner. 
Typically, full-reference image quality assessment measures are aggregated over the entire image using a global mean. However, line charts often occupy only a small part of the canvas, and aggregating them using a global mean would result in smoothened-out metrics influenced by the canvas's filling level. To address this issue and facilitate cross-template comparisons, we introduce OR-conv masking. This technique aggregates only the per-pixel values (in the neighborhood) of rendered data points in the visualization. 
For further details, we refer to a description containing examples and implementation details in \href{https://github.com/predict-idlab/ts-datapoint-selection-vis/blob/main/details/OR-conv_masking.md}{the GitHub repository}.

\begin{figure*}[!tb]
 \centering 
 \includegraphics[width=\linewidth]{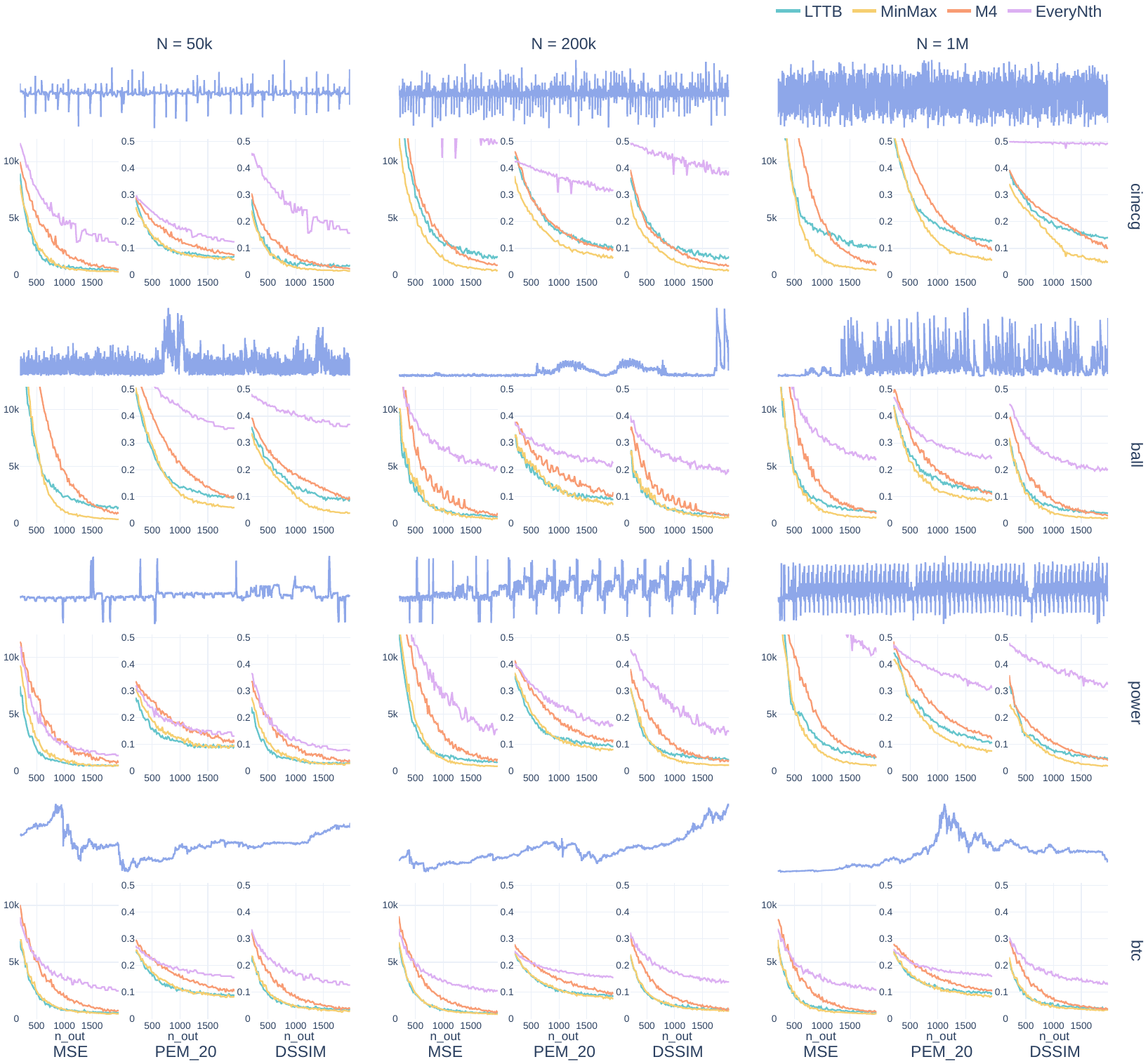}
 \caption{Assessing visual representativeness of data point selection algorithms for various image templates. Each row displays a distinct time series dataset, with columns indicating the template size. All image templates, including the blue reference templates which are depicted above the metric subplots, were generated using Plotly's default settings (linear interpolation, line width of 2). The metric subplots reveal trends in aggregation method performance as $n_{out}$ (x-axis) increases (range [200, 2000]). $PEM\_20$ refers to Pixel Error with a Margin of 20, where pixels with a difference above 20 are binarized and divided by the conv-mask size for a ratio. $DSSIM$ denotes OR-conv mask scaled structural dissimilarity (see \cref{sec:dssim}), while MSE represents conv-mask scaled Mean Squared Error.}


 \vspace*{-3mm}
 \label{fig:vis_repr_grid}
\end{figure*}

\subsection{Visual Stability}
Visual stability evaluates visual changes that occur when updating the displayed view with limited new data. 
The stability of downsampling algorithms is a crucial aspect to consider, given that interactivity is essential for effective exploration of time series data~\cite{van2022plotly}, as also highlighted by Shneiderman's visual information seeking mantra: \textit{"Overview first, zoom and filter, then details-on-demand"}~\cite{shneiderman_eyes_nodate}.

Visual stability assessment consists of two primary cases: panning and zooming. 
Panning involves shifting the displayed data range by an offset $\Delta \neq 0$, resulting in a range of $[x_{start} + \Delta, x_{end} + \Delta]$. Positive offsets correspond to right panning, while negative offsets signify left panning. 
Zooming changes the data range to $[x_{start} + \Delta_l, x_{end} + \Delta_r]$ with at least one $\Delta \neq 0$ and $\Delta_l \neq \Delta_r$ (as $\Delta_l = \Delta_r$ is panning). Zooming can be further differentiated into two subcases: zooming in ($\Delta_l \geq 0$, $\Delta_r \leq 0$) and zooming out ($\Delta_l \leq 0$, $\Delta_r \geq 0$). A positive left offset denotes left zooming in, a negative right offset corresponds to right zooming in, while their combination represents left-and-right zooming in. Conversely, positive right and negative left offsets indicate respectively right and left zooming out, and their combination implies left-and-right zooming out. Remark that streaming corresponds to zooming out with $\Delta_l = 0$ and $\Delta_r > 0$. Additionally, it is worth noting that visual stability is complementary to visual representativeness, as the latter addresses larger zooms (due to varying template sizes $N$), while the former concentrates on smaller updates.

Visual stability is primarily influenced by the chosen downsampling algorithm rather than the visualization toolkit's specific configuration. Although an image-space approach could assess stability, it faces challenges such as the need for perfectly overlapping pixels and the inability to change the y-range when new global extrema are added during an update. To avoid these limitations and effectively quantify stability, we utilize data-space metrics. Our approach involves first interpolating the updated downsampled data to the original downsampled data at their intersection, enabling data-space comparisons at the same x-values. We then calculate two stability metrics: the Mean Absolute Error (MAE) and the Mean Absolute Error at the Peaks (MAEP). Note that these measures assess stability with respect to the original downsampled data and do not consider the visual representativeness of the downsampled time series, which is measured by visual representativeness metrics.


\subsubsection{Mean Absolute Error (MAE)}
The MAE is a measure that quantifies the average difference between the original downsampled data and the updated downsampled data. To make the absolute errors more comparable across different templates, we normalize them by the min-max range of the original downsampled data. 
This measure takes all downsampled data points into account, providing an overall indication of the absolute differences. However, as a direct consequence, this metric is susceptible to smoothing out a few large errors.

\subsubsection{Mean Absolute Error at Peaks (MAEP)}
The MAEP is calculated in a similar manner as the MAE explained above, but only considers peaks of the data. Peaks are defined as the local maxima or minima in the original aggregated data and are more visually prominent than other parts of the time series~\cite{kim2021towards}. This metric is partly in accordance with the max pendicular distance proposed by Shi et al.~\cite{shi2006performance}, which aims to capture the largest deviation. This makes MAEP a more sensitive measure of visual stability than the MAE, as it only focuses on the most visually prominent parts of the data.

\section{Results}
In this section we will present and discuss our findings when applying the proposed methodology considering the EveryNth, MinMax, LTTB, and M4 algorithms. The two subsections delve further into the two aspects of our methodology.

\begin{figure*}[!t]
 \centering
 \includegraphics[width=\linewidth]{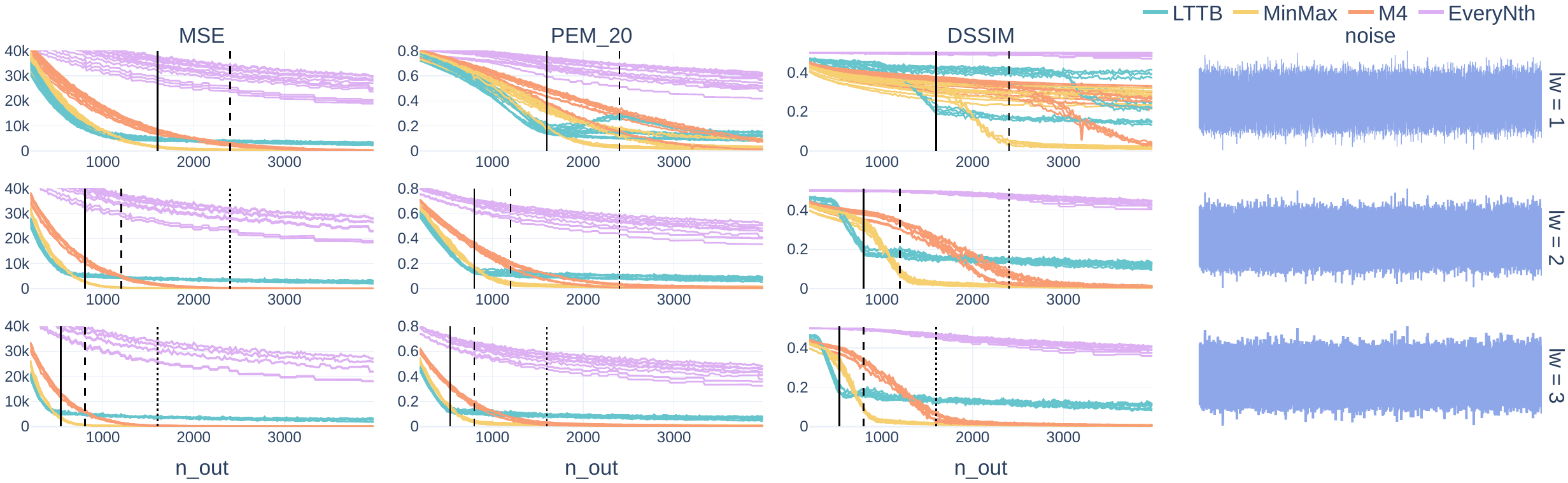}
 \vspace*{-5mm}
 \caption{Visual representativeness of noise for varying line widths (1, 2, 3), data sizes (50k, 200k, 1M), and toolkits (plotly, bokeh, matplotlib), with $n_{out}$ 200-4000. The aggregation images are compared to reference images of the same line width.}
 \vspace*{-3mm}
 \label{fig:noise_line_width}
\end{figure*}

\subsection{Visual Representativeness}\label{sec:evaluation_vis_representativity}
We assess visual representativeness using three widely used Python visualization toolkits: Matplotlib, Plotly, and Bokeh. These toolkits were chosen because they are the most prominent visualization tools in Python, which is the leading language for interoperating with large data~\cite{noauthor_kaggles_2022}. Specifically, we examine the visual representativeness of data point selection algorithms through two approaches.

First, we maintain a constant line-chart configuration while altering image templates, focusing on Plotly's default settings. This experimental design mirrors the typical user behavior of not adjusting these settings. Through this analysis, we can identify trends in downsampling algorithms concerning data efficiency across various templates, without the interference of varying configurations.

Second, we utilize a fixed time series template, specifically a stationary noise signal, to investigate the effects of different visualization libraries and line widths on the visual representativeness of data point selection algorithms.

The final subsection explores the particularities of implementing a pixel-perfect M4 aggregation algorithm in practice.

\subsubsection{Data Efficiency \& General Trends}
\cref{fig:vis_repr_grid} presents the visual representativeness metrics for Plotly's default configuration across four data point selection algorithms. By visualizing the metrics over $n_{out}$ (x-axis of each metric-subplot), we can evaluate data efficiency, which is the performance of these algorithms at specific (low) $n_{out}$ values. A similar visualization using Plotly's default configuration for the two other toolkits can be found in the code repository~\footnote{\label{gh-readme-gifs}\href{https://github.com/predict-idlab/ts-datapoint-selection-vis/blob/main/details/vis_representativity.md}{github:ts-datapoint-selection-vis/details/visual-representativity}}.

A general observation reveals that the trend and order of the aggregators remain relatively consistent across metrics for each template subplot. As a second, anticipated trend, we note that increasing $n_{out}$ enhances visual representativeness. Regarding the order of aggregators, we observe that MinMax and LTTB (depending on the template) generally perform the best, followed by M4 in third place, and EveryNth performing the worst. Additionally, the LTTB and MinMax curves tend to display similar shapes within each metric, with the elbow occurring in the same place, albeit with a slightly different slope. These observations hold true for the other two toolkits as well, see Footnote 3.

In terms of data efficiency, MinMax and LTTB stand out as the most data-efficient aggregators, yielding the best metric values for $n_{out} < 1000$. Depending on the template, one may slightly outperform the other. These findings are consistent with those reported by Gil et al.~\cite{gil2021towards}, whose results identified LTTB as a data-efficient algorithm. 
MinMax excels with rough templates such as cinecg-200k, cinecg-1M, ball-1M, and power-1M. For all other templates, we observe identical performance for $n_{out} < 800$. As $n_{out}$ continues to increase, we notice that MinMax converges to LTTB's performance or lower. Particularly for high-roughness image templates (e.g., ball-50k, cinecg-1M), MinMax converges to lower metric values than LTTB (and M4). 
For low-roughness data, such as all bitcoin templates, LTTB and MinMax demonstrate identical performance even for larger $n_{out}$.
M4 and EveryNth exhibit poorer data efficiency and converge considerably slower than MinMax and LTTB, which aligns with expectations, given EveryNth's naive selection and M4 outputting 4 selected data points per bin, reducing its data efficiency.

The PEM\_20 metric converges to a value of 0.1 for most image templates, which can be attributed to shading errors caused by anti-aliasing~\footnote{This \href{https://github.com/predict-idlab/ts-datapoint-selection-vis/blob/main/gifs/toolkit_aa.gif}{GIF} illustrates the effect of aliasing on the PEM\_20.}. This observation is further supported by the fact that MSE decreases to near-zero values for large $n_{out}$, indicating that smaller (shading) errors are the primary contributors to the 0.1 PEM\_20 value. Remark that as the PEM\_20 is scaled with the OR conv-mask size, its value decreases more for image templates that fill up more canvas space, where less relative space is occupied by shading at the outer edges of the filled-up space, e.g., ball-50k, cinecg-1M, and power-1M.

\subsubsection{Data Efficiency \& Line Width}\label{sec:evaluation_vis_repr_stationary_noise}
\cref{fig:vis_repr_grid} presents the visual representativeness metrics for the stationary noise time series signal, where each row corresponds to a distinct line width. These visualizations include all three toolkits, and the template size $N$ is varied within the subplot, as it has a minimal impact on the image template for stationary noise data.

A key insight is the consistent trend of the aggregator performance curves across various data sizes and toolkits for each of the metric subplots. Remark that some inconsistencies occur in the DSSIM and PEM\_20 subplots for line width 1, for which we provide a detailed explanation on the GitHub repository, see Footnote 3.

LTTB emerges as the most data-efficient approach for this time series signal, which can be attributed to its maximization of triangular surfaces. This algorithm is particularly effective at covering large areas due to its preference for alternating between bin-wise minimum and maximum values. In the teaser \cref{fig:teaser}, this alternation between minimum and maximum values is clearly visible for the LTTB aggregation of noisy data. 
Nonetheless, LTTB does not converge to the lowest value, which can also be described by the algorithm's preference for minimum-maximum alternation, since this can lead to the exclusion of global extrema that are chosen by MinMax and M4. Consequently, M4 will ultimately converge to the same values as MinMax, as can clearly be observed in the $lw = 3$ row. Additionally, it is worth noting that M4 and EveryNth tend to converge at a considerably slower rate.

For each of the aggregators, we notice an elbow in the performance curves, i.e., where the curve converges, for the PEM\_20 and DSSIM metrics. We notice that the $n_{out}$ position of this elbow decreases when increasing the line width. Indicating that increasing line width can be a means to improve data efficiency.
Upon further investigation, we derived a general formula to identify the specific $n_{out}$ position of the observed elbow as a function of canvas width ($cw$) and line width ($lw$): 
                    $$elbow = \frac{cw * \bar{n}_{ga}}{lw}$$
Here, $\bar{n}_{ga}$ represents the number of selected data points necessary to have a \textit{guaranteed alternation}. Since LTTB favors min-max alternation, $\bar{n}_{ga}$ is $2$. In contrast, MinMax can require up to 3 data points to achieve alternation, resulting in an $\bar{n}_{ga}$ of $3$. M4 can be considered as a variation of MinMax that also selects the first and last data points in a bucket, and therefore requires twice the $\bar{n}_{ga}$ of MinMax, i.e., $6$. \cref{fig:noise_line_width} depicts the elbow positions for $\bar{n}_{ga}$ values of 2 (LTTB), 3 (MinMax), and 6 (M4) using solid, dashed, and dotted vertical lines, respectively.
The trend for higher data efficiency with increasing line width is also apparent for the image template grid \footnote{This \href{https://github.com/predict-idlab/ts-datapoint-selection-vis/blob/main/gifs/plotly_default_slider\%3Dlw.gif}{GIF} illustrates the effect of line width on the image template grid.}.

\subsubsection{Pixel-perfect M4}
The previous subsections focused on data efficiency for visual representativeness, which is not a strong selling point for M4, but its pixel-perfectness is. As noted by Jugel et al., achieving pixel-perfect M4 requires a $n_{out}$ that is four times the pixel-width of the canvas~\cite{jugel_m4_2014}. This subsection explores the practicalities necessary to create a pixel-perfect aggregation with the M4 algorithm.

\begin{figure}[!tb]
 \centering
    \includegraphics[width=\linewidth]{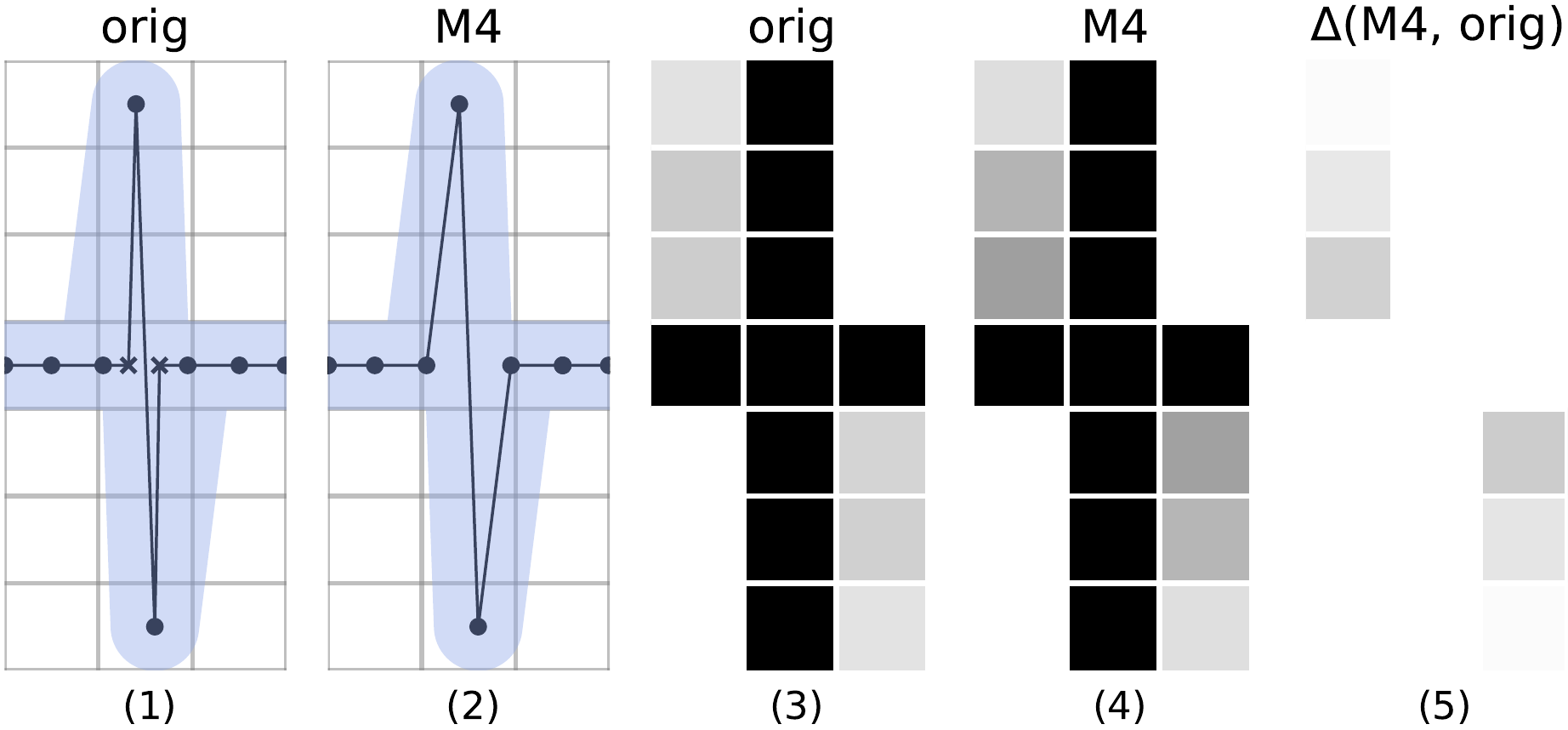}
    \vspace*{-5mm}
    \caption{Visualization of M4-related pixel errors when anti-aliasing is applied. Subplot (1) presents the original time series featuring a 1-pixel wide shading. Upon applying M4 in subplot (2), the two data points marked by an $\times$-symbol in (1), are excluded. Subplots (3) and (4) depict the anti-aliased renderings of the original and M4 aggregated time series, respectively, while subplot (5) emphasizes the differences between them.}
    \label{fig:aa_m4}
    \vspace{-3mm}
\end{figure}

Our first focus is on anti-aliasing, also known as pixel binarization, which is a common technique employed to enhance the perceptual quality of visualizations~\cite{leler1980human}. In all three toolkits investigated in this study, anti-aliasing is enabled by default. Moreover, Bokeh and Plotly do not offer the option to render aliased figures (disabling anti-aliasing). A straightforward and effective approach to anti-aliasing, suitable for simple graphics like line charts, is to determine the percentage of pixel occupancy for a given region in the vector graphic and use that percentage as color shading. \cref{fig:aa_m4} applies this approach to illustrate that M4 fails to produce pixel-perfect visualizations when anti-aliasing is employed. Subplot (2) indicates how removing data points with M4 results in minor deviations in the slope and starting point of the shaded line, causing differences in the coloring of anti-aliased pixels due to varying percentages of pixel area occupancy (5). Remark that subplots (3) and (4) were constructed using Matplotlib's default rasterization backend, demonstrating the direct effect of M4 aggregation for this toolkit.

Another aspect of visualization configuration influencing M4 its pixel-perfectness is line width. Increasing the line width leads to larger shaded areas, potentially causing greater pixel shading errors. Bokeh's default line width is set to 1 pixel, Plotly's default is 2 pixels, and Matplotlib defines line width in points, corresponding to $1/72$ of the figure's DPI (measured in inches). Consequently, achieving pixel-perfect M4 without manually adjusting line widths in these toolkits proves to be challenging.

Aside from configuring aliasing and line width, we observed that achieving pixel-perfect M4 requires modifying other visualization-related parameters, such as rasterization back-end and x-range binning.

Finally, to establish a bin to pixel-column mapping ($n_{out} = 4 * cw$), one must either control or be informed of the canvas width in advance, which is not always trivial. Graph layout elements, including y-axis ticks, legends, and font sizes, influence the eventual canvas width. Additionally, these elements can change due to updates, such as y-axis ticks changing following a user graph interaction, further complicating canvas width assessment. Furthermore, web-based visualization frameworks, such as Plotly and Bokeh can stretch across the entire webpage and are susceptible to browser page resizing and zooming. 

In conclusion, achieving pixel-perfect visualization with M4 is difficult or even impossible in most existing visualization toolkits, given the necessity for substantial control over the visualization environment. 
Jugel et al. mentioned that "in a real implementation, the engineers have to make sure that $n_{out} = 4 * cw$ to achieve the best results"~\cite{jugel_m4_2014}, but this statement was only a minor point in their paper. Bae et al. presented a practical implementation of M4 in D3, yet they neither mentioned anti-aliasing nor verified the pixel-perfectness of their implementation~\cite{bae2017practical}. 
As demonstrated above, there are more components which influence M4's performance than initially anticipated. Therefore, we argue that M4 should not be considered merely as another data point selection algorithm, but rather as a data-efficient visualization technique, given its tight coupling to the rendering of data points.

\subsection{Visual Stability}

\begin{figure}[!t]
    \centering
    \includegraphics[width=\linewidth]{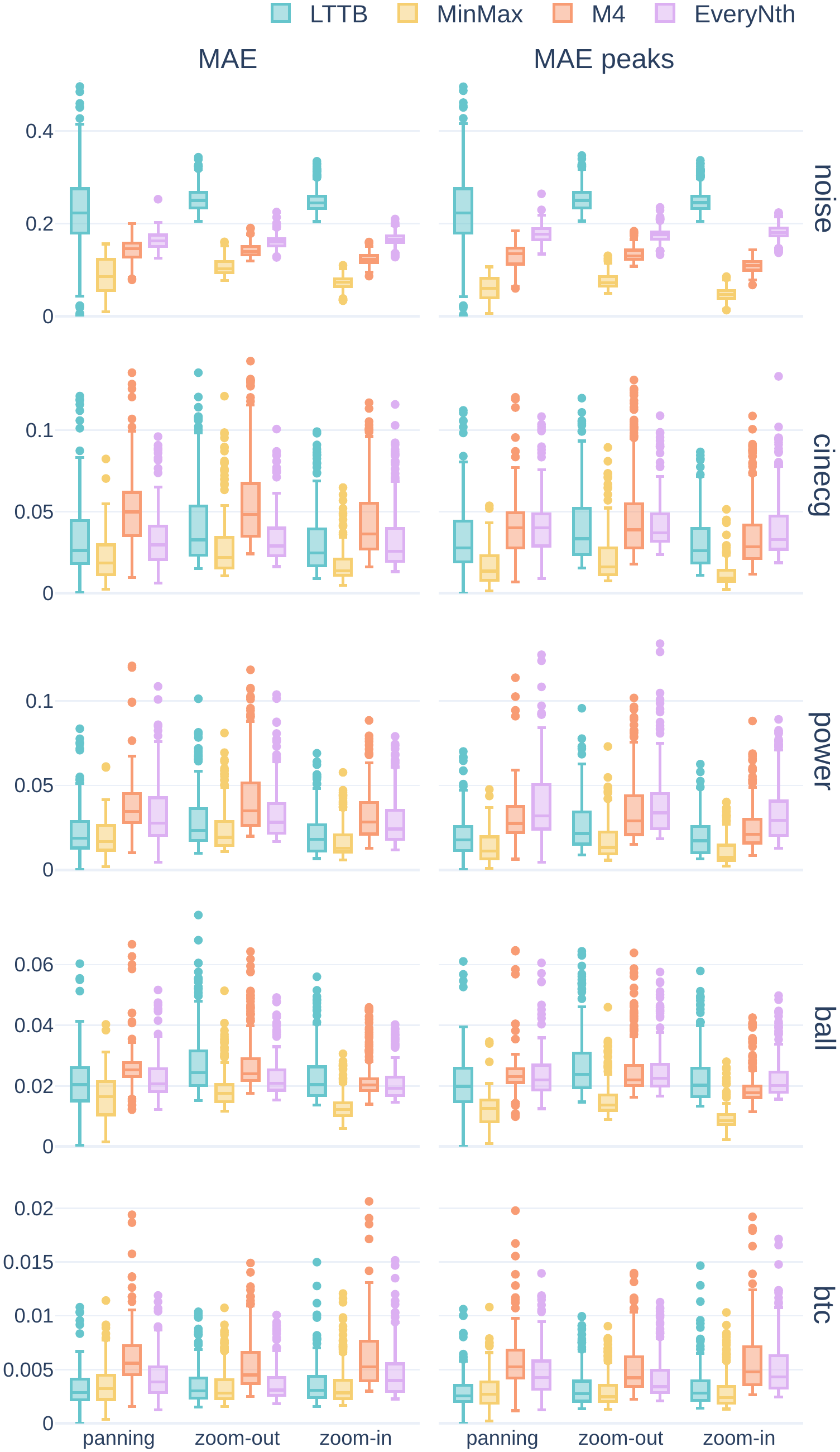}    
    \caption{Assessment of visual stability for various data point selection algorithms applied to different time series templates. Each column represents a unique time series template with a length of 200K. The two rows display the $MAE$ and $MAEP$ distributions for the examined downsampling algorithms, distinguished by color. Each box-plot account for multiple offsets (in both left and right directions) and include a range of $n_{out}$ values from 200 to 2000, in increments of 200.}
    \label{fig:visual_stability}
    \vspace{-3mm}
\end{figure}

As detailed in the methodology section, visual stability is categorized into two primary cases: panning and zooming. Zooming can be further differentiated into zooming in and zooming out (the latter corresponds, to some extent, to streaming new data). Consequently, we analyze three distinct cases: panning, zooming in, and zooming out. The offsets, defined as ratios, are calculated based on $N$ and include the following values: $[1/53, 1/27, 1/17, 1/11, 3/23, 2/11]$. Note that these selected offset ratios all have a prime number in the denominator, which corresponds more to real-world user interactions instead of aligned offsets. For instance, selecting 0.1 as the offset ratio would result in perfectly aligned bins over the intersection, which will yield 0 MAE(P) measures. 




\cref{fig:visual_stability} shows visual stability results for five datasets, each row representing a time series template of length 200k. We focused on the 200K templates to avoid excessively low or high roughness found in 50K and 1M templates, see \cref{fig:template_grid}. Additionally, incorporating noise inherently covers high roughness cases.
For the Bitcoin time series in the last row, MinMax and LTTB demonstrated superior stability compared to the other algorithms, with M4 being the most unstable. 
In the subsequent three rows above, which correspond to the ball, power, and cinecg datasets, MinMax consistently surpasses the stability of the other algorithms, often by a substatial margin. 
When examining the $MAE$, M4 frequently performs the worst, followed by either EveryNth or LTTB. Regarding the $MEAP$, EveryNth demonstrates to be the most unstable, followed by M4 and then LTTB. Notably, when focusing solely on the peaks ($MEAP$), the margin between MinMax and the other algorithms became even more pronounced. 
For the generated noise in the last column, MinMax once again emerges as the most stable algorithm, while LTTB, in contrast to previous templates, is the least stable.
Another noteworthy observation is the minimal variation in the results across the three different cases. Additionally, this \href{https://github.com/predict-idlab/ts-datapoint-selection-vis/blob/main/gifs/visual_stability_n_out.gif}{GIF} supplements our findings by illustrating how increasing $n_{out}$ contributes to improved stability, a logical yet interesting result\footnote{Further details regarding visual stability, showing that at least 30\% of the data points are included for the MEAP metric, can be found in the \href{https://github.com/predict-idlab/ts-datapoint-selection-vis/blob/main/details/vis_stability.md}{GitHub repository}.}.

A plausible explanation for MinMax's superior stability lies in its high \textit{locality}, referring to the extent that data point selection depends on individual bin values without influence from neighboring bins. High locality results in greater stability and reduced alignment issues susceptibility.
MinMax has the highest locality among all considered algorithms, as it selects the minimum and maximum values for each bin with no interplay of neighboring bins. 
In contrast, LTTB exhibits lower locality, initializing its sequential data point selection process with the first data point and then using the previously selected data point from the preceding bin (together with the average point of the next bin) as input for the local algorithm. Additionally, the local algorithm chooses the point that maximizes the triangular surface, which, as described in \cref{sec:evaluation_vis_repr_stationary_noise}, causes LTTB to favor bin-wise minimum-maximum alternation, further amplifying the propagation of alignment issues throughout the data point selection process. This interplay makes LTTB more susceptible to alignment issues, as evident in \cref{fig:visual_stability} with generated noise. 
In particular, for high-noise data, the instability of LTTB is most prominent, as there are many local peaks to choose from.
Furthermore, the EveryNth algorithm is even more prone to alignment, as the offset directly impacts all selected data points.
M4, a combination of MinMax and EveryNth, results in half of the chosen data points being susceptible to alignment. In M4, the minimum and maximum values are always between (or at) the first and last points, causing two out of three visually interpolated lines per bin to be sensitive to the offset (only the interpolation between the minimum and maximum values remains largely unaffected). Consequently, we observe that M4 is most of the time on par with or slightly performs better than EveryNth, but not achieving the stability provided by MinMax. This is particularly noticeable in the case of generated noise, see \cref{fig:visual_stability} and an illustration in the teaser \cref{fig:teaser}.

\section{Guidelines}
Based on the evaluation results, we formulate the following evidence-based guidelines for time series visualization at scale with data point selection:

\begin{enumerate}

\item \textbf{Use MinMax or LTTB for data-efficient aggregation}: MinMax and LTTB have shown to be the most data-efficient aggregators for low $n_{out}$. However, for larger $n_{out}$, MinMax often outperforms LTTB's visual representativeness. 

\item \textbf{Use MinMax for superior visual stability}: MinMax has consistently exhibited superior visual stability, outperforming other algorithms significantly. While LTTB ranked second in most cases, it exhibited high instability when numerous peaks are present, stemming from LTTB favoring bin-wise extrema alternation. 

\item \textbf{Consider pixel-perfect M4 as a visualization technique rather than a data point selection algorithm}: Achieving pixel-perfect M4 requires aliased figures and control over various visualization-related parameters, such as line width, rasterization back-end, x-range binning, and canvas width. 

\item \textbf{Consider line width when aiming for high data efficiency}: Increasing line width has proven to be an effective means for improving data efficiency at the cost of granularity. We identified a clear relationship between line width, canvas width, and $n_{out}$, which defines the elbow of the performance curve.

\end{enumerate}

While MinMax seems to have received only little attention, for instance, the studies conducted by Gil et al.\cite{gil2021towards}, Bae et al.~\cite{bae2017practical}, and  Steinarsson~\cite{steinarsson_downsampling_nodate} did not consider this algorithm, our results nevertheless indicate that MinMax is a highly competitive downsampling algorithm, often outperforming other algorithms. Only Jugel et al.~\cite{jugel_m4_2014}, in accordance with our results, acknowledged the higher data efficiency of MinMax as a (minor) observation of their evaluation. 
Therefore, we argue that MinMax deserves more attention in future studies. 

\section{Conclusion}
This work presents an extensive metrics-based methodology for evaluating data point selection algorithms for visualization, quantifying visual representativeness and a novel concept called "visual stability. Our methodology employs various time series datasets, each of three different sizes, accounting for a wide range of time series properties and zoom levels. 
To measure visual representativeness, we used three image-based metrics, allowing us to assess the influence of visualization toolkit parameters. We facilitated consistent comparisons across different time series templates, by introducing an OR-conv mask for more targeted per-pixel metric aggregation. Furthermore, we advocate using the number of selected data points as a more fair measure of data efficiency.
Visual stability is measured using two data-space metrics, overcoming rasterization alignment limitations that would present themselves when evaluating in image space.


We evaluated the EveryNth, MinMax, M4, and LTTB algorithms using our proposed methodology, taking various figure-drawing properties into account. 
In most cases, LTTB and/or MinMax emerged as superior algorithms in terms of visual representativeness. 
We also explored the impact of line width on data efficiency and derived a general formula, based on our empirical results, using line width, canvas width, and the number of guaranteed alternations $\bar{n}_{ga}$, that defines the elbow of the performance curve.
Furthermore, we showed that realizing pixel-perfect M4 is nontrivial or even impossible in practice. Notably, we are the first to observe that M4 can never be pixel-perfect in the presence of anti-aliasing.
As for visual stability, MinMax proved to be the most stable algorithm, often by a significant margin. We provided an explanation for these results by relating to the locality of downsampling algorithms. 
Based on these insights, we derived a set of guidelines for large-scale time series visualization with data point selection. A general conclusion is that the MinMax algorithm deserves more attention, given its superior performance and not being considered in most related studies.

The proposed methodology and obtained results are publicly available, enabling reproducibility and further research in this domain. This study provides a foundation for obtaining new insights and assessing downsampling algorithms.

\subsection{Limitations}
Although this study provides numerous insights on time series data point selection for visualization, some limitations should be acknowledged. 
Firstly, we scoped this study, in line with other related studies~\cite{steinarsson_downsampling_nodate, rong_asap_2017, jugel_m4_2014, gil2021towards}, to regularly sampled data without any gaps, which may not be fully representative of real-world time series data. 
Additionally, no user studies were employed, and while metrics can capture visual representativeness and visual stability to some degree, they are proxies for user perception. Therefore, future work could consider employing user studies to identify thresholds for the utilized metrics below which the perception does not improve. Our openly available code base provides a sound foundation for such studies, as two of the investigated toolkits are web-integrable, enabling web-based surveys.
%
Finally, this study only investigated a limited set of visualization configurations that influence visual representativeness, such as line width and toolkit. Line shape and other configurations were not investigated, as these are less frequently changed, but the proposed methodology and experiments allow for investigating these in future studies.

\section*{Author Contributions}
\textbf{Jonas Van Der Donckt}: Conceptualization, Methodology, Software, Validation, Formal Analysis, Data Curation, Writing - Original Draft, Visualization.
\textbf{Jeroen Van Der Donckt}: Conceptualization, Methodology, Software, Validation, Formal Analysis, Data Curation, Writing - Original Draft, Visualization.
\textbf{Michael Rademaker}: Review \& Editing.
\textbf{Sofie Van Hoecke}: Review \& Editing.

\section*{Supplemental Materials}
\label{sec:supplemental_materials}

The supplemental materials are made available on GitHub at \url{https://github.com/predict-idlab/ts-datapoint-selection-vis}.
This repository includes (1) an implementation of the four investigated downsampling algorithms, (2) code for reproducing the conducted experiments, (3) parquet files containing the collected data, and (4) preprocessing of the utilized data together with a reference to where one can download this data.






\acknowledgments{%
    The authors wish to thank Louise Van Calenbergh for proofreading the manuscript.
    Jonas Van Der Donckt (1S56322N) is funded by a doctoral fellowship of the Research Foundation Flanders (FWO).
    Part of this work is done in the scope of the imec.AAA Context-aware health monitoring project. 
}

\bibliographystyle{abbrv}

\bibliography{template}

\begin{thebibliography}{10}

\bibitem{agrawal_challenges_2015}
R.~Agrawal, A.~Kadadi, X.~Dai, and F.~Andres.
\newblock Challenges and opportunities with big data visualization.
\newblock In {\em Proceedings of the 7th {International} {Conference} on
  {Management} of computational and collective {intElligence} in {Digital}
  {EcoSystems}}, pages 169--173, Caraguatatuba Brazil, Oct. 2015. ACM.

\bibitem{aigner_visualizing_2007}
W.~Aigner, S.~Miksch, W.~Müller, H.~Schumann, and C.~Tominski.
\newblock Visualizing time-oriented data—{A} systematic view.
\newblock {\em Computers \& Graphics}, 31(3):401--409, June 2007.
\newblock ZSCC: 0000450.

\bibitem{bae2017practical}
P.~Bae, K.-W. Lim, W.-S. Jung, and Y.-B. Ko.
\newblock Practical implementation of m4 for web visualization service.
\newblock {\em Journal of Communications and Networks}, 19(4):384--391, 2017.

\bibitem{datashaderpitfalls}
J.~A. Bednar and J.~Signell.
\newblock Common plotting pitfalls that get worse with large data.
\newblock
  \href{https://github.com/holoviz/datashader/blob/a36adab573907be0fccf46161b93a6e3eebc403e/examples/user_guide/1_Plotting_Pitfalls.ipynb}{\nolinkurl{github.com/holoviz/datashader/examples/user\_guide/1\_Plotting\_Pitfalls.ipynb}}.

\bibitem{bikakis_big_2018}
N.~Bikakis.
\newblock Big {Data} {Visualization} {Tools}.
\newblock {\em arXiv:1801.08336 [cs]}, Feb. 2018.
\newblock ZSCC: 0000075 arXiv: 1801.08336.

\bibitem{breddels_interactive_2016}
M.~A. Breddels.
\newblock Interactive (statistical) visualisation and exploration of a billion
  objects with vaex.
\newblock {\em Proceedings of the International Astronomical Union},
  12(S325):299--304, Oct. 2016.

\bibitem{dau2019ucr}
H.~A. Dau, A.~Bagnall, K.~Kamgar, C.-C.~M. Yeh, Y.~Zhu, S.~Gharghabi, C.~A.
  Ratanamahatana, and E.~Keogh.
\newblock The ucr time series archive.
\newblock {\em IEEE/CAA Journal of Automatica Sinica}, 6(6):1293--1305, 2019.

\bibitem{timescaleblog_paganini_2023}
J.~Davi~Paganini.
\newblock Downsampling in the database: How data locality can improve data
  analysis.
\newblock
  \href{https://www.timescale.com/blog/downsampling-in-the-database-how-data-locality-can-improve-data-analysis/}{\nolinkurl{www.timescale.com/blog}},
  Feb 2023.

\bibitem{douglas1973algorithms}
D.~H. Douglas and T.~K. Peucker.
\newblock Algorithms for the reduction of the number of points required to
  represent a digitized line or its caricature.
\newblock {\em Cartographica: the international journal for geographic
  information and geovisualization}, 10(2):112--122, 1973.

\bibitem{franco2022multivariate}
J.~Franco, A.~Garcia, and A.~Gil.
\newblock Multivariate adaptive downsampling algorithm for industry 4.0 data
  visualization.
\newblock In {\em 16th International Conference on Soft Computing Models in
  Industrial and Environmental Applications (SOCO 2021)}, pages 588--597.
  Springer, 2022.

\bibitem{gil2021towards}
A.~Gil, M.~Quartulli, I.~G. Olaizola, and B.~Sierra.
\newblock Towards smart data selection from time series using statistical
  methods.
\newblock {\em IEEE Access}, 9:44390--44401, 2021.

\bibitem{gorodov_analytical_2013}
E.~Y. Gorodov and V.~V. Gubarev.
\newblock Analytical {Review} of {Data} {Visualization} {Methods} in
  {Application} to {Big} {Data}.
\newblock {\em Journal of Electrical and Computer Engineering}, 2013:1--7,
  2013.

\bibitem{hellerstein1999control}
J.~M. Hellerstein, R.~Avnur, A.~Chou, C.~Hidber, C.~Olston, V.~Raman, T.~Roth,
  and P.~J. Haas.
\newblock Interactive data analysis: The control project.
\newblock {\em Computer}, 32(8):51--59, 1999.

\bibitem{datashader}
{Holoviz}-community.
\newblock Datashader, quickly and accurately render even the largest data.
\newblock \url{https://github.com/holoviz/datashader}.

\bibitem{jerzak2012debs}
Z.~Jerzak, T.~Heinze, M.~Fehr, D.~Gr{\"o}ber, R.~Hartung, and N.~Stojanovic.
\newblock The debs 2012 grand challenge.
\newblock In {\em Proceedings of the 6th ACM International Conference on
  Distributed Event-Based Systems}, pages 393--398, 2012.

\bibitem{jugel_m4_2014}
U.~Jugel, Z.~Jerzak, G.~Hackenbroich, and V.~Markl.
\newblock M4: a visualization-oriented time series data aggregation.
\newblock {\em Proceedings of the VLDB Endowment}, 7(10):797--808, June 2014.

\bibitem{jugel_vdda_2016}
U.~Jugel, Z.~Jerzak, G.~Hackenbroich, and V.~Markl.
\newblock {VDDA}: automatic visualization-driven data aggregation in relational
  databases.
\newblock {\em The VLDB Journal}, 25:53--77, Feb. 2016.

\bibitem{noauthor_kaggles_2022}
Kaggle-inc.
\newblock Kaggle's {State} of {Machine} {Learning} and {Data} {Science} 2022.
\newblock Technical report, Oct. 2022.

\bibitem{kim2021towards}
D.~H. Kim, V.~Setlur, and M.~Agrawala.
\newblock Towards understanding how readers integrate charts and captions: A
  case study with line charts.
\newblock In {\em Proceedings of the 2021 CHI Conference on Human Factors in
  Computing Systems}, pages 1--11, 2021.

\bibitem{kottarathil_btc_data_2022}
P.~Kottarathil.
\newblock Bitcoin historical dataset.
\newblock
  \href{https://www.kaggle.com/datasets/prasoonkottarathil/btcinusd}{\nolinkurl{kaggle.com/datasets/btcinusd}},
  Mar 2022.

\bibitem{kwon_sampling_2017}
B.~C. Kwon, J.~Verma, P.~J. Haas, and C.~Demiralp.
\newblock Sampling for {Scalable} {Visual} {Analytics}.
\newblock {\em IEEE Computer Graphics and Applications}, 37(1):100--108, Jan.
  2017.

\bibitem{leler1980human}
W.~J. Leler.
\newblock Human vision, anti-aliasing, and the cheap 4000 line display.
\newblock {\em ACM Siggraph Computer Graphics}, 14(3):308--313, 1980.

\bibitem{lin_visualizing_2005}
J.~Lin, E.~Keogh, and S.~Lonardi.
\newblock Visualizing and {Discovering} {Non}-{Trivial} {Patterns} in {Large}
  {Time} {Series} {Databases}.
\newblock {\em Information Visualization}, 4(2):61--82, June 2005.
\newblock ZSCC: 0000178.

\bibitem{mutschler2013debs}
C.~Mutschler, H.~Ziekow, and Z.~Jerzak.
\newblock The debs 2013 grand challenge.
\newblock In {\em Proceedings of the 7th ACM international conference on
  Distributed event-based systems}, pages 289--294, 2013.

\bibitem{nilsson2020understanding}
J.~Nilsson and T.~Akenine-M{\"o}ller.
\newblock Understanding ssim.
\newblock {\em arXiv preprint arXiv:2006.13846}, 2020.

\bibitem{uber_raskin_aggarwal_2018}
B.~Raskin and N.~Aggarwal.
\newblock The billion data point challenge: Building a query engine for high
  cardinality time series data.
\newblock
  \href{https://www.uber.com/en-BE/blog/billion-data-point-challenge/}{\nolinkurl{uber.com/billion-data-point-challenge}},
  Dec 2018.

\bibitem{rong_asap_2017}
K.~Rong and P.~Bailis.
\newblock {ASAP}: prioritizing attention via time series smoothing.
\newblock {\em Proceedings of the VLDB Endowment}, 10(11):1358--1369, Aug.
  2017.

\bibitem{sara2019image}
U.~Sara, M.~Akter, and M.~S. Uddin.
\newblock Image quality assessment through fsim, ssim, mse and psnr—a
  comparative study.
\newblock {\em Journal of Computer and Communications}, 7(3):8--18, 2019.

\bibitem{shi2006performance}
W.~Shi and C.~Cheung.
\newblock Performance evaluation of line simplification algorithms for vector
  generalization.
\newblock {\em the cartographic journal}, 43(1):27--44, 2006.

\bibitem{shneiderman_eyes_nodate}
B.~Shneiderman.
\newblock The {Eyes} {Have} {It}: {A} {Task} by {Data} {Type} {Taxonomy} for
  {Information} {Visualizations}.
\newblock In {\em Proceedings 1996 IEEE symposium on visual languages}, pages
  336--343. IEEE, 1996.

\bibitem{steinarsson_downsampling_nodate}
S.~Steinarsson.
\newblock Downsampling {Time} {Series} for {Visual} {Representation}.
\newblock Master's thesis, University of Iceland, 2013.

\bibitem{van2022plotly}
J.~Van Der~Donckt, J.~Van Der~Donckt, E.~Deprost, and S.~Van~Hoecke.
\newblock Plotly-resampler: Effective visual analytics for large time series.
\newblock In {\em 2022 IEEE Visualization and Visual Analytics (VIS)}, pages
  21--25. IEEE, 2022.

\bibitem{visvalingam1993line}
M.~Visvalingam and J.~D. Whyatt.
\newblock Line generalisation by repeated elimination of points.
\newblock {\em The cartographic journal}, 30(1):46--51, 1993.

\bibitem{walker_timenotes_2016}
J.~Walker, R.~Borgo, and M.~W. Jones.
\newblock {TimeNotes}: {A} {Study} on {Effective} {Chart} {Visualization} and
  {Interaction} {Techniques} for {Time}-{Series} {Data}.
\newblock {\em IEEE Transactions on Visualization and Computer Graphics},
  22(1):549--558, Jan. 2016.

\bibitem{wang2009mean}
Z.~Wang and A.~C. Bovik.
\newblock Mean squared error: Love it or leave it? a new look at signal
  fidelity measures.
\newblock {\em IEEE signal processing magazine}, 26(1):98--117, 2009.

\bibitem{zhai2020perceptualIQA}
G.~Zhai and X.~Min.
\newblock Perceptual image quality assessment: a survey.
\newblock {\em Science China Information Sciences}, 63:1--52, 2020.

\end{thebibliography}

\newpage
\onecolumn

\end{document}